\documentclass[12pt,a4paper]{article}

\usepackage{graphicx}
\usepackage{color}
\usepackage{epsfig}
\usepackage{amsmath,amssymb,bm}
\usepackage{booktabs}
\usepackage{url}

\usepackage[utf8]{inputenc}
\usepackage[numbers,sort&compress]{natbib}
\usepackage[colorlinks,
            linkcolor=blue,
            filecolor=blue,
            anchorcolor=blue,
            urlcolor=blue,
            citecolor=blue
            ]{hyperref}

\usepackage{bookmark}

\numberwithin{equation}{section}

\makeatletter
\def\Slash{\mathpalette\@Slash}
\def\@Slash#1#2{{\ooalign{\hfil$#1/$\hfil\crcr$#1{#2}$}}}
\makeatother

\allowdisplaybreaks

\begin{document}

\title{\bf \Large R-parity violating solutions to the \boldmath{$R_{D^{(\ast)}}$} anomaly and their GUT-scale unifications}

\author{
  \centerline{
    Quan-Yi Hu\footnote{qyhu@mails.ccnu.edu.cn},
    Xin-Qiang Li\footnote{xqli@mail.ccnu.edu.cn},
    Yu Muramatsu\footnote{yumura@mail.ccnu.edu.cn},~and
    Ya-Dong Yang\footnote{yangyd@mail.ccnu.edu.cn}}
  \\[25pt]
  \centerline{
    \begin{minipage}{\linewidth}
      \begin{center}
{\it \normalsize
    Institute~of~Particle~Physics~and~Key~Laboratory~of~Quark~and~Lepton~
    Physics~(MOE), Central~China~Normal~University,~Wuhan,~Hubei~430079,~
    People's~Republic~of~China}\\[3pt]
      \end{center}
    \end{minipage}}
  \\[10pt]}

\date{}
\maketitle
\vspace{0.2cm}

\begin{abstract}
\noindent Recently, several $B$-physics experiments report interesting anomalies in the semi-leptonic decays of $B$-mesons, such as the excess in the $R_{D^{(\ast)}}$ measurements.
These anomalies seem to suggest intriguing hints of lepton flavor non-universality, and the R-parity violating (RPV) interactions are candidates for explaining this non-universality.
In this paper, we discuss the RPV interactions for resolving the $R_{D^{(\ast)}}$ anomaly with the Grand Unified Theory (GUT) assumption. To solve the $R_{D^{(\ast)}}$ anomaly, it is known that large RPV couplings and around $1~{\rm TeV}$ sfermion masses are required. At the same time, large RPV couplings are conducive to realize the bottom-tau Yukawa unification which appears in the GUT models. On the other hand, there are problems for realizing favorable sfermion masses in the constrained minimal supersymetric standard model. To resolve these problems, we show that two non-universalities, the non-universal sfermion masses and the non-universal gaugino masses, are favorable.
\end{abstract}

\newpage

\section{Introduction \label{sec.1}}

Recently, several interesting anomalies in semi-leptonic $B$-meson decays are reported by the $B$-physics experiments. For example, the ratios of the branching fractions
\begin{equation}
R_{D^{(\ast)}}=\frac{\mathcal{B}(B \to D^{(\ast)} \tau \nu)}{\mathcal{B}(B \to D^{(\ast)} l \nu)},
\end{equation}
where $l=e,\,\mu$, have been measured by the BaBar~\cite{BaBar}, Belle~\cite{Belle}, and LHCb~\cite{LHCb} collaborations and show large deviations from the Standard Model (SM) predictions. These ratios are ideal observables for testing the SM, especially the lepton flavor universality, because they are independent of the Cabibbo-Kobayashi-Maskawa (CKM) matrix element $V_{cb}$ and are free from the experimental and theoretical uncertainties common to the numerator and the denominator. The most recent averages of the BaBar, Belle, and LHCb measurements, as compiled by the Heavy Flavor Averaging Group~\cite{Amhis:2016xyh}, read~\cite{Amhis:2018up}
\begin{equation}
R_D^{\text{avg}}=0.407 \pm 0.039 \pm 0.024,\quad
R_{D^\ast}^{\text{avg}}=0.306 \pm 0.013 \pm 0.007,
\end{equation}
exceeding the SM predictions~\cite{Amhis:2018up,Bigi:2016mdz, Bernlochner:2017jka, Bigi:2017jbd, Jaiswal:2017rve}
\begin{equation}
R_D^\text{SM}=0.299 \pm 0.003,\quad
R_{D^\ast}^{\text{SM}}=0.258 \pm 0.005,
\end{equation}
by $2.3\sigma$ and $3.0\sigma$, respectively. Considering the $R_D^{\text{avg}}-R_{D^\ast}^{\text{avg}}$ correlation of $-0.203$, the resulting combined difference from the SM predictions would be about $3.78\sigma$~\cite{Amhis:2018up}. These excesses, if confirmed with more precise experimental data and theoretical predictions, would suggest the presence of lepton flavor non-universality. Therefore, to understand this non-universality, many studies have been done both within the model-independent frameworks~\cite{Freytsis:2015qca,RD_model-indep} as well as in some specific New Physics models, such as leptoquarks~\cite{Freytsis:2015qca,LQ}, charged Higgses~\cite{Freytsis:2015qca,HPM}, and charged vector bosons~\cite{Freytsis:2015qca,WPM}; for reviews on this excess see, for example, Ref.~\cite{review_RD}.

As demonstrated in Refs.~\cite{Altmannshofer:2017poe,RPV_RD}, the R-parity violating (RPV) supersymetric (SUSY) models~\cite{Hall:1983id,Barbier:2004ez} are also viable candidates for explaining the $R_{D^{(\ast)}}$ anomaly.
In the minimal supersymetric SM (MSSM) with RPV interactions, an exchange of right-handed down-type squarks coupled to quarks and leptons can yield the required four-fermion interactions contributing to $R_{D^{(\ast)}}$ at the tree level, which is similar to the case with leptoquark induced interactions. In Ref.~\cite{Altmannshofer:2017poe}, the RPV contributions to $R_{D^{(\ast)}}$ are discussed within a minimal effective RPV-SUSY scenario, and it is shown that $\mathcal{O}(1)$ RPV coupling and around $1~{\rm TeV}$ sbottom mass are required to satisfy the $R_{D^{(\ast)}}$ measurements, while being consistent with other experimental constraints as well as preserving the gauge coupling unification. In this work, we shall discuss this possibility within the Grand Unified Theory (GUT) framework.

GUTs~\cite{Georgi:1974sy,GUT_review} are interesting candidates for physics beyond the SM. In the GUTs, two interesting unifications, the gauge coupling unification and the matter unification, are realized. The matter unification plays an important role in restricting the model parameters. In the MSSM~\cite{Martin:1997ns}, there are enormous SUSY breaking parameters and, in order to make these SUSY breaking parameters restrictive, the constrained MSSM (CMSSM) is introduced as a very well-motivated, realistic and concise SUSY extension of the SM. In the CMSSM~\cite{Kane:1993td}, the matter unification unifies the SUSY breaking parameters; for example, gaugino masses are unified into one universal gaugino mass. Moreover, for obtaining a realistic GUT framework, the introduction of RPV interactions is beneficial~\cite{Dreiner:1995hu}. In the $SU(5)$ GUTs, the down and charged-lepton Yukawa couplings are unified, which is however not easy to be realized in the standard minimal setup. The RPV interactions modify the renormalization group (RG) flow for realizing this unification~\cite{Dreiner:1995hu} and, as we will show in this paper, these RPV couplings are also large enough to explain the $R_{D^{(\ast)}}$ anomaly.

In this paper, we will examine the sfermion masses within the GUT framework. As we will show later, models with the CMSSM assumption do not satisfy the $R_{D^{(\ast)}}$ measurements, because a large RPV coupling makes these models already excluded, and the right-handed sbottom mass becomes too heavy to explain the $R_{D^{(\ast)}}$ anomaly. To solve this problem, two non-universalities, the non-universal sfermion masses and the non-universal gaugino masses, which are compatible with the GUT framework, play an important role. Moreover, these non-universalities are motivated to stabilize the electro-weak (EW) scale and will be future signals of our scenario.

This paper is organized as follows. In Sec. \ref{sec.2}, we explain the relations among the RPV interactions, the $R_{D^{(\ast)}}$ anomaly, and the $SU(5)$ unification. In Sec. \ref{sec.3}, we first show the problem encountered with the CMSSM assumption for resolving the $R_{D^{(\ast)}}$ anomaly, and then demonstrate the necessary two non-universalities for figuring out this problem. Our discussion and summary are given in Sec. \ref{sec.4}. For convenience, we give all the relevant formulae for the processes used to produce Figure \ref{fig:RPV_para_region} in Appendix A, while the density plots for the sfermion masses at  1 TeV are shown in Appendix B.

\section{RPV, \texorpdfstring{\boldmath{$R_{D^{(\ast)}}$}}{Lg} anomaly, and \texorpdfstring{\boldmath{$SU(5)$}}{Lg} unification \label{sec.2}}

First of all, we specify the RPV couplings in the MSSM. The most general renormalizable superpotential consistent with the gauge symmetry and field content of the MSSM is given by~\cite{Weinberg:1981wj,Allanach:2003eb}
\begin{align}\label{eq:potential}
W&=\epsilon_{ab}\Big[ (Y_e)_{ij} L_i^a H_d^b E_j^c + (Y_d)_{ij} Q_i^{a\alpha} H_d^b D_{j\alpha}^c + (Y_u)_{ij} Q_i^{a\alpha} H_u^b U_{j\alpha}^c \Big] - \epsilon_{ab} \Big[ \mu H_d^a H_u^b + \mu_i L_i^a H_u^b \Big] \nonumber\\[0.2cm]
&+\epsilon_{ab} \Big[ \tfrac{1}{2} (\Lambda_{e_k})_{ij} L_i^a L_j^b E_k^c + (\Lambda_{d_k})_{ij} L_i^a Q_j^{b\alpha} D_{k\alpha}^c \Big] + \tfrac{1}{2} \epsilon_{\alpha \beta \gamma} (\Lambda_{u_i})_{jk} U_i^{c\alpha} D_j^{c\beta} D_k^{c\gamma},
\end{align}
where $L$, $Q$, $E^c$, $D^c$, $U^c$, $H_d$, and $H_u$ are the chiral superfields for the MSSM multiplet, and we denote the $SU(3)_C$ and $SU(2)_L$ fundamental representation indices by $\alpha, \beta, \gamma = 1, 2, 3$ and $a, b = 1, 2$, respectively, while the generation indices by $i, j, k = 1, 2, 3$.  $\epsilon_{ab}$ and $\epsilon_{\alpha \beta \gamma}$, with $\epsilon_{12}=\epsilon_{123}=+1$, are the totally anti-symmetric tensors for the $SU(2)_L$ and $SU(3)_C$ gauge groups, respectively.

In addition to the R-parity conserving (RPC) couplings $(Y_e)_{ij}$, $(Y_d)_{ij}$, $(Y_u)_{ij}$ and $\mu$, the following RPV couplings are introduced in Eq.~\eqref{eq:potential}: lepton number violating tri-linear couplings $(\Lambda_{e_k})_{ij}$ and $(\Lambda_{d_k})_{ij}$, baryon number violating tri-linear couplings $(\Lambda_{u_i})_{jk}$, and lepton number violating bi-linear couplings $\mu_i$. As the lepton doublet superfields $L_i$ and the Higgs doublet superfield $H_d$ have the same gauge and Lorentz quantum numbers in MSSM, the $\mu_i$ term in Eq.~\eqref{eq:potential} can be sent to zero via a rotation in the $(L_i, H_d)$ space~\cite{Hall:1983id,Banks:1995by}, which will be assumed throughout this paper.

\subsection{RPV interactions contributing to \texorpdfstring{\boldmath{$R_{D^{(\ast)}}$}}{Lg} \label{sec.1.1}}

As the underlying quark-level transition in $R_{D^{(\ast)}}$ involve quarks and leptons, at the tree level we need only consider the $\Lambda_d$ term in Eq.~\eqref{eq:potential}. Working in the mass eigenstates for the down-type quarks and assuming sfermions are in their mass eigenstates, one  obtains from Eq.~\eqref{eq:potential} the following effective Lagrangian contributing to the transition $d_n \to u_j e_i \nu_m$ at the tree-level after integrating out the heavy squarks~\cite{Altmannshofer:2017poe,RPV_RD}:
\begin{equation} \label{eq:Leff}
\mathcal{L}_{\text{eff}} \supset -\frac{(\Lambda_{d_k})_{ij}(\Lambda_{d_k})_{mn}^\ast}{2 m^2_{\tilde{d}_{k}}} \left[ \bar{\nu}_{Lm} \gamma^\mu e_{Li} \bar{d}_{Ln} \gamma_\mu (V^\dagger_{CKM}u_L)_j \right] + {\rm h.c.},
\end{equation}
where $V_{CKM}$ is the CKM matrix and $m_{\tilde{d}_{k}}$ denotes the mass of the right-handed squark. Specifying to the decays $B\to D^{(\ast)} \ell \nu$ ($\ell=e,\, \mu$, or $\tau$), one can see that the resulting four-fermion operator has the same chirality structure as in the SM, implying that the RPV contribution to $R_{D^{(\ast)}}$ is simply a rescaling of the SM result.

\begin{figure}[h!t]
\centering
\includegraphics[width=0.95\textwidth,pagebox=cropbox,clip]{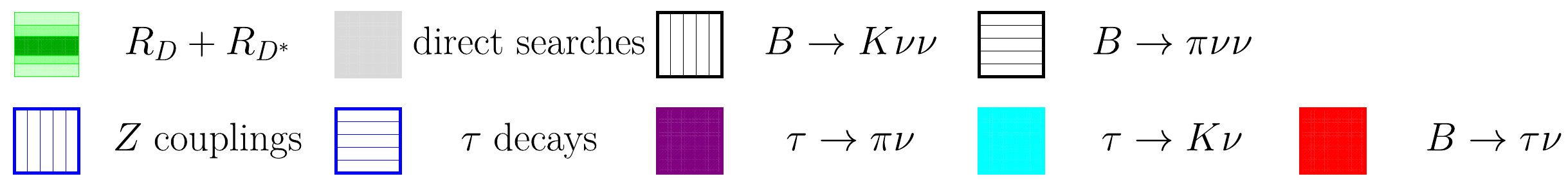}\\[0.2cm]
\includegraphics[width=0.95\textwidth,pagebox=cropbox,clip]{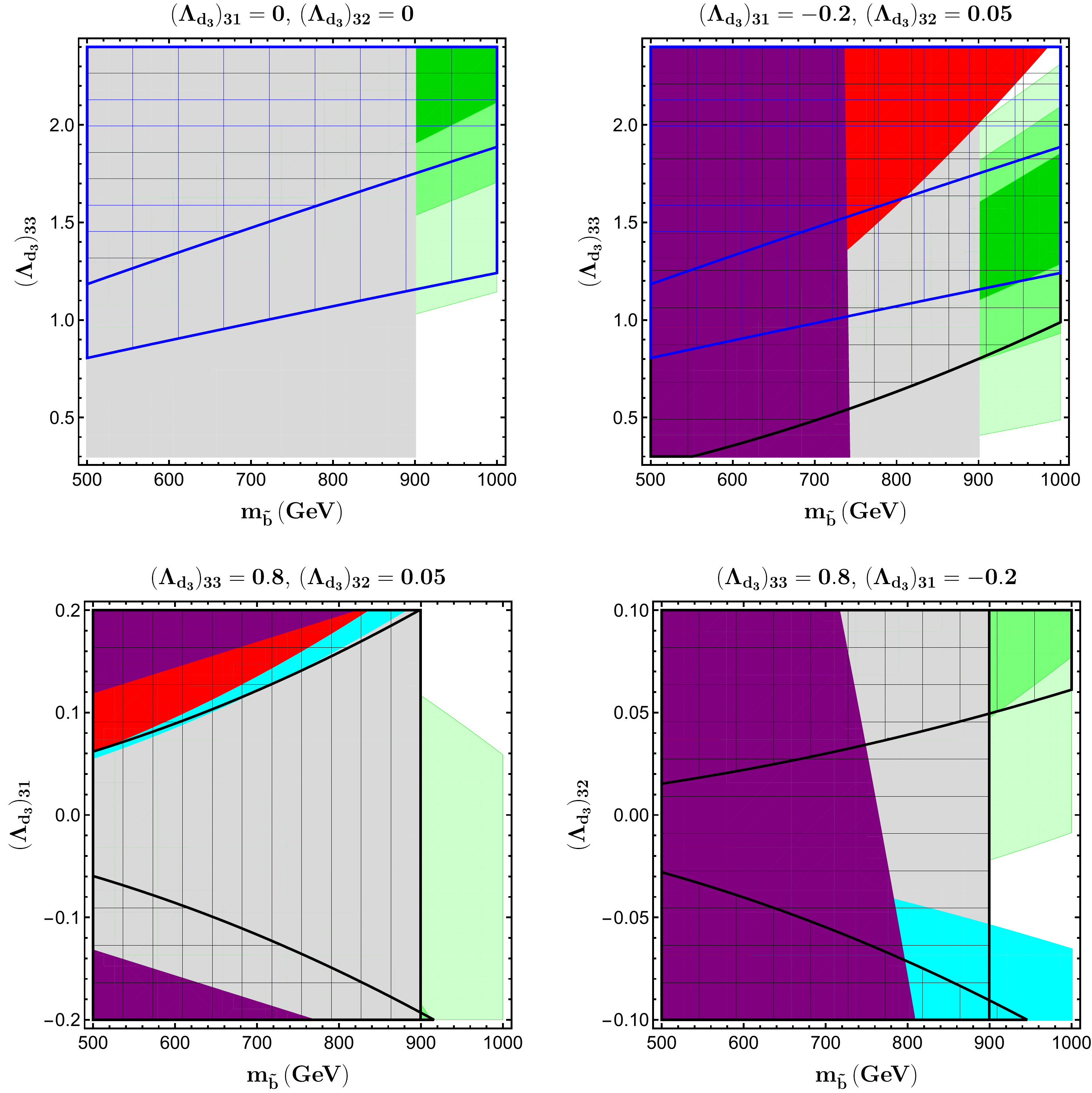}
\caption{\small Parameter region satisfying the $R_{D^{(\ast)}}$ measurements as well as the other relevant constraints. The SM predictions and the experimental data are summarized in Table \ref{tab:const_diff}.}
\label{fig:RPV_para_region}
\end{figure}

The contribution of the effective operator in Eq.~\eqref{eq:Leff} to the $R_{D^{(\ast)}}$ anomaly has been discussed in Refs.~\cite{Altmannshofer:2017poe,RPV_RD}, and it is shown that large values of $(\Lambda_{d_3})_{33}\sim1-2$ are required to explain the $R_{D^{(\ast)}}$ anomaly within $1\sigma$ level, but such a large $(\Lambda_{d_3})_{33}$ coupling develops a Landau pole below the GUT scale~\cite{Altmannshofer:2017poe}. Along the same line as in Ref.~\cite{Altmannshofer:2017poe}, we show in Figure \ref{fig:RPV_para_region} the parameter region in the $(m_{\tilde{b}}, (\Lambda_{d_3})_{3j})$ plane that satisfy the $R_{D^{(\ast)}}$ measurements as well as the constraints from other relevant processes. For convenience, we have also given in Appendix \ref{app.eq} all the relevant formulae for the processes used to produce Figure \ref{fig:RPV_para_region}.

\begin{table}[t]
\tabcolsep 0.05in
\renewcommand\arraystretch{1.3}
\begin{center}
\caption{\small SM predictions and experimental data for $R_{D^{(\ast)}}$ and the other relevant observables. Differences between the experimental data used in Ref.~\cite{Altmannshofer:2017poe} and here are also shown.}
\vspace{0.1cm}
{\footnotesize
\begin{tabular}{|c|c|c|c|}
\hline\hline
 & & Ref.~\cite{Altmannshofer:2017poe} & our work\\ \cline{3-4}
 & SM prediction & experimental result & experimental result \\
\hline
$R_D$ & $0.299 \pm 0.003$~\cite{Amhis:2018up,Bigi:2016mdz, Bernlochner:2017jka,Jaiswal:2017rve} & $0.403 \pm 0.040 \pm 0.024$~\cite{Amhis:2016up}  & $0.407 \pm 0.039 \pm 0.024$~\cite{Amhis:2018up}\\
$R_{D^\ast}$ & $0.258 \pm 0.005$~\cite{Amhis:2018up, Bernlochner:2017jka, Bigi:2017jbd, Jaiswal:2017rve} & $0.310 \pm 0.015 \pm 0.008$~\cite{Amhis:2016up} & $0.306 \pm 0.013 \pm 0.007$~\cite{Amhis:2018up}\\
$\mathcal{B}(B \to \tau \nu) \times 10^{4}$ & $0.947 \pm 0.182$~\cite{Nandi:2016wlp}  & $1.06 \pm 0.19$~\cite{BtaunuNov2016} & $1.44 \pm 0.31$~\cite{BtaunuDec2017}  \\
$\mathcal{B}(B^+ \to K^+ \nu \nu) \times 10^{6}$ & $3.98 \pm 0.43 \pm 0.19$~\cite{Buras:2014fpa}  & $<16$~\cite{Lees:2013kla} & same  \\
$\mathcal{B}(B^+ \to \pi^+ \nu \nu) \times 10^{7}$ & $1.46 \pm 0.14$~\cite{Du:2015tda}  & $<980$~\cite{Lutz:2013ftz} & same  \\
$\mathcal{B}(\tau \to \pi \nu)$ & $(10.91\pm0.024)\%$  & none & $(10.82\pm0.05)\%$~\cite{Patrignani:2016xqp} \\
$\mathcal{B}(\tau \to K \nu) \times 10^3$ & $7.15\pm0.026$  & none & $6.96\pm0.10$~\cite{Patrignani:2016xqp} \\
$Z$ couplings: $\frac{g_{Z\tau_L\tau_L}}{g_{Zl_Ll_L}}$ & 1  & $1.0013 \pm 0.0019$~\cite{ALEPH:2005ab} & same  \\
$\tau$ decays: $\frac{g_{W\tau_L\nu_\tau}}{g_{Wl_L\nu_l}}$ & 1  & $1.0007 \pm 0.0013$~\cite{Amhis:2016xyh,Altmannshofer:2017poe} & same  \\
direct searches: $m_{\tilde{b}}$ &   & $> 680~{\rm GeV}$~\cite{Khachatryan:2015bsa} & $> 900~{\rm GeV}$~\cite{Sirunyan:2018nkj}  \\
\hline \hline
\end{tabular}
}
\label{tab:const_diff}
\end{center}
\end{table}

In our calculation, we have used the latest measurements as well as the updated SM predictions which are summarized in Table \ref{tab:const_diff}. In this table, we also summarize differences between the experimental data used in Ref.~\cite{Altmannshofer:2017poe} and here. To obtain the allowed parameter region, we use the following best fit value in the RPV scenario:
\begin{equation}
\frac{R_D}{R_D^{\text{SM}}}=\frac{R_{D^\ast}}{R_{D^\ast}^{\text{SM}}}=1.22 \pm 0.05.
\end{equation}
In Figure \ref{fig:RPV_para_region}, the $1\sigma$-, $2\sigma$-, and $3\sigma$-favored regions from the $R_{D^{(\ast)}}$ measurements are shown in dark-green, green, and light-green, respectively. It is clearly seen, especially from the top-right plot in Figure \ref{fig:RPV_para_region}, that a realization of the $R_{D^{(\ast)}}$ measurements at the $2\sigma$ level is possible. We have also considered the extra constraints from $D$ and $\tau$ decays as discussed in Ref.~\cite{Bhattacharyya:1995pq}, and found that only some $\tau$-decay modes, especially the decay $\tau \to K \nu$, can provide complementary constraints on the parameter regions allowed by the $R_{D^{(*)}}$ measurements, while the leptonic charm decays $D\to\tau\nu$ and $D_s\to\tau\nu$ do not lead to any relevant constraints in our scenario.

In Ref.~\cite{Altmannshofer:2017poe}, the $R_{D^{(\ast)}}$ anomaly was discussed in a minimal effective SUSY scenario with RPV.
In such a natural SUSY scenario, the masses of the third-generation sfermions are below $1~{\rm TeV}$, while the first- and second-generation sfermions are quite heavy and can be thought of being decoupled from the low-energy spectrum~\cite{Dimopoulos:1995mi}. This mass spectrum explains the current experimental results naturally, because the decoupled first- and second-generation sfermions suppress the SUSY flavor contributions, and the light third-generation sfermion masses support stabilizing the EW scale. At the same time, such a natural mass spectrum also realizes large RPV couplings~\cite{Brust:2011tb}.

\subsection{RPV couplings and GUT-scale unification \label{sec.1.2}}

From the last subsection, we have seen that $\mathcal{O}(1)$ RPV couplings for sbottom masses compatible with the current direct searches are helpful for realizing the $R_{D^{(\ast)}}$ measurements. However, the RPV couplings are strongly constrained by the proton stability~\cite{Barbier:2004ez}: the conservative limit for the combination of the lepton and baryon number violating couplings is $\left| \Lambda_d \Lambda_u \right| < 10^{-11}$ for any generation indices~\cite{Smirnov:1996bg}. Basically, in SUSY $SU(5)$ GUT models with RPV, all the tri-linear RPV couplings are induced from a single term $\Lambda_{ijk} {\bf 10}_i {\bf \bar{5}}_j {\bf \bar{5}}_k$, and should therefore preserve the following relation at the GUT scale~\cite{Georgi:1974sy}:
\begin{equation}
\Lambda_e = \Lambda_d = \Lambda_u.
\end{equation}
To satisfy the conservative limit set by the proton stability, either the RPV couplings become negligibly small or some novel mechanism should be invoked to evade this limit~\cite{Smirnov:1995ey,Brahm:1989iy}. The proton stability constraint is extremely severe when both the lepton and baryon number violating RPV couplings are simultaneously non-zero. For simplicity, we implicitly assume that the proton stability is ensured by some mechanism\footnote{There are many mechanisms that can be used to prevent rapid proton decay~\cite{Smirnov:1995ey,Brahm:1989iy}. In the matter-Higgs mixing mechanism~\cite{Smirnov:1995ey}, for example, the only source of RPV couplings comes from the mixing between the down-type Higgs doublet and the left-handed lepton doublet and, therefore, only the lepton number violating RPV couplings are generated and could be in principle of the order 1. }, and focus only on the $\Lambda_{d_3}$ couplings which are crucial for accommodating the $R_{D^{(\ast)}}$ anomaly.

In $SU(5)$ GUT models, the gauge couplings and the Yukawa matrices satisfy their respective unification relations~\cite{Georgi:1974sy}:
\begin{equation}
g_1=g_2=g_3,\quad Y_e = Y_d^T,
\end{equation}
with $i=1,\,2,\,3$ for the $U(1)_Y$, $SU(2)_L$, and $SU(3)_C$ gauge groups, respectively. Here the hypercharge gauge coupling $g_1$ includes already the GUT normalization factor. Among the unification relations of Yukawa matrices, the one for the third generation, the bottom-tau Yukawa unification $y_{\tau}=y_b$, is particularly promising, because small Yukawa couplings for the first and second generations can be modified at the GUT scale through contributions from higher-dimensional operators~\cite{Ellis:1979fg} and/or higher-dimensional representations of Higgs fields~\cite{Georgi:1979df}. It is known that the unification relation for the third generation can be satisfied thanks to the RPV contributions~\cite{Dreiner:1995hu}.

\begin{table}[t]
\tabcolsep 0.25in
\renewcommand\arraystretch{1.3}
\begin{center}
\caption{\small Input parameters involved in our calculation for the RG flow of the gauge, Yukawa and RPV couplings. Only central values from Ref.~\cite{Patrignani:2016xqp} are used.}
\vspace{0.1cm}
\begin{tabular}{|c|c||c|c|}
\hline\hline
$\alpha$ & $1/137.036$ & $m_b (m_b)$ & $4.18~{\rm GeV}$ \\
$\alpha_s(M_Z)$ & $0.1182$ & $M_t$ (direct measurements) & $173.1~{\rm GeV}$ \\
$\sin^2\theta_{W}$ & $0.23129$ & $M_\tau$ & $1776.86~{\rm MeV}$ \\
$M_Z$ & $91.1876~{\rm GeV}$ &  & \\
\hline\hline
\end{tabular}
\label{tab:input}
\end{center}
\end{table}

To see the effect of these RPV contributions, we calculate the RG flow for the gauge, Yukawa and RPV couplings. The relevant input parameters involved in our calculation are summarized in Table \ref{tab:input}. These parameters at the GUT scale are obtained by the following four steps: First, we use the Mathematica package \texttt{RunDec}~\cite{Chetyrkin:2000yt} for the running and converting of different quark masses, and follow Ref.~\cite{Arason:1991ic} to translate the lepton pole masses into the running masses, for calculating the $\overline{\text{MS}}$ Yukawa couplings at the $M_Z$ scale. Second, we calculate these parameters at $1~{\rm TeV}$ from the SM RG flow at the two-loop level~\cite{Luo:2002ey}. Third, we follow Ref.~\cite{Martin:1993yx} to translate the $\overline{\text{MS}}$ scheme parameters into the $\overline{\text{DR}}$ scheme ones. Finally, we evaluate these parameters at the GUT scale with the presence of RPV couplings, using the two-loop RG flow for the RPC parameters~\cite{Martin:1993zk} and the one-loop RG flow for the RPV parameters~\cite{Allanach:2003eb}. In this paper, the GUT scale $M_{\text{GUT}}$ is defined by
\begin{equation}
g_1 (M_{\text{GUT}}) = g_2 (M_{\text{GUT}}).
\end{equation}

\begin{figure}[t]
  \centering
  \includegraphics[width=0.46\textwidth,pagebox=cropbox,clip]{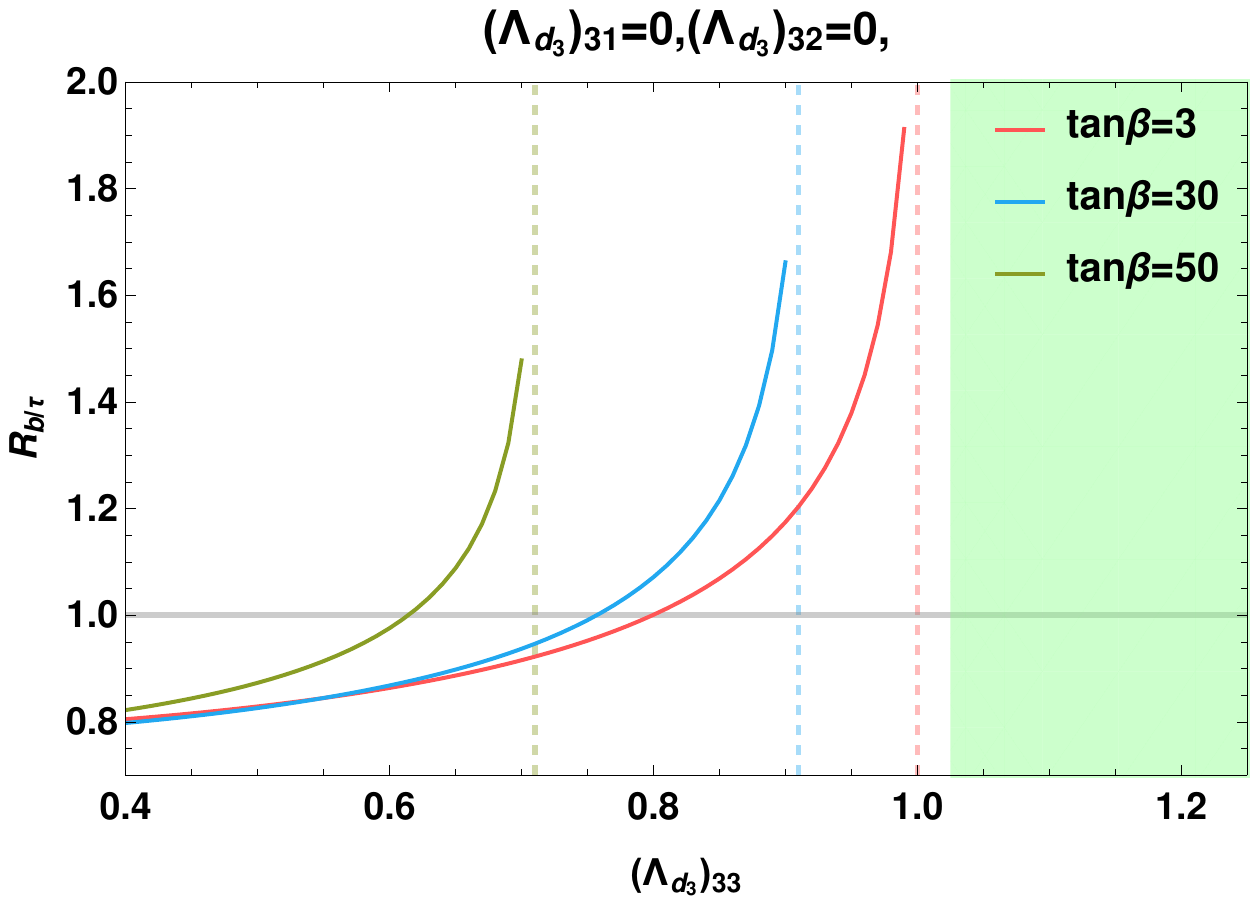}\qquad
  \includegraphics[width=0.46\textwidth,pagebox=cropbox,clip]{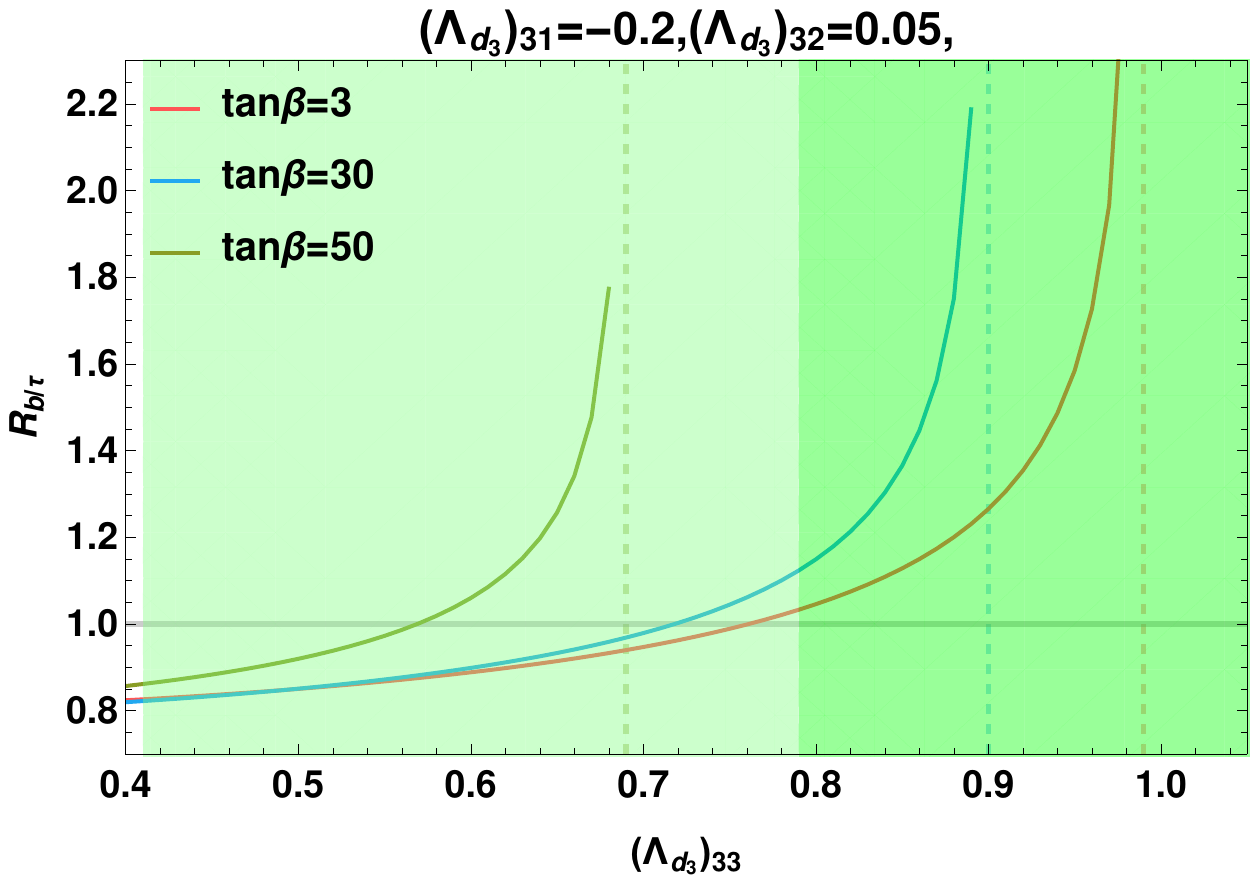}
  \caption{\small Dependence of the ratio $R_{b/\tau}\equiv y_b/y_\tau$ on the RPV coupling $(\Lambda_{d_3})_{33}$ for three representative values of $\tan \beta$. The favored region by the $R_{D^{(\ast)}}$ measurements is shown in green. Upper bounds for escaping the Landau pole before the GUT scale are shown by the dashed lines.}
\label{fig:Rbtau}
\end{figure}

In Figure \ref{fig:Rbtau}, the contribution of the RPV coupling $(\Lambda_{d_3})_{33}$ to $R_{b/\tau}\equiv y_b/y_\tau$ is shown for three representative values of $\tan \beta$, the ratio of the Higgs vacuum expectation values (VEVs). The favored region for $(\Lambda_{d_3})_{33}$ by the $R_{D^{(\ast)}}$ measurements is shown in green when $m_{\tilde{b}}= 900~{\rm  GeV}$. In the left panel of Figure \ref{fig:Rbtau}, we make all the RPV couplings except for $(\Lambda_{d_3})_{33}$ zero and, in the right panel, we take $(\Lambda_{d_3})_{31}=-0.2$ and $(\Lambda_{d_3})_{32}=0.05$. The dashed line shows the upper bound for escaping the Landau pole before the GUT scale. From this figure, the following two points can be made: first, to realize the bottom-tau Yukawa unification, large $(\Lambda_{d_3})_{33}$ is desirable; second, both the ratio $R_{b/\tau}$ and the Landau pole depend strongly on the value of $\tan \beta$. This dependence is due to a $\tan\beta$ dependence of the Yukawa couplings $y_b$ and $y_\tau$~\cite{Allanach:2003eb}: large $\tan\beta$ makes $y_b$ and $y_\tau$ large through the matching condition of the Yukawa couplings between the SM and the MSSM, and large $y_b$ and $y_\tau$ move the Landau pole for the RPV coupling $(\Lambda_{d_3})_{33}$ closer to the low-energy scale. Therefore, to solve the $R_{D^{(\ast)}}$ anomaly, small $\tan \beta$ is particularly favorable.

\section{Sfermion masses needed for \texorpdfstring{\boldmath{$R_{D^{(\ast)}}$}}{Lg} anomaly \label{sec.3}}

Up to now, we have shown that $\mathcal{O}(1)$ RPV couplings are helpful for resolving the $R_{D^{(\ast)}}$ anomaly when the third-generation sfermion masses are around $1~{\rm TeV}$, and $\mathcal{O}(1)$ RPV couplings are also favorable for realizing the $SU(5)$ Yukawa relation, the bottom-tau Yukawa unification at the GUT scale. In this section, we proceed to discuss the impact of large RPV couplings on the sfermion masses.

In the beginning, we calculate these contributions in the CMSSM scenario~\cite{Kane:1993td}. In this scenario, the SUSY parameters at the GUT scale are defined as
\begin{equation}
M_1=M_2=M_3=m_{1/2},
\end{equation}
\begin{equation}
(m_{\tilde{q}}^2)_{ij}=(m_{\tilde{l}}^2)_{ij}=(m_{\tilde{u}}^2)_{ij}=(m_{\tilde{d}}^2)_{ij}=(m_{\tilde{e}}^2)_{ij}=m_0^2 \mathbf{1}_{ij},\quad
m_{h_u}^2=m_{h_d}^2=m_0^2,
\end{equation}
\begin{equation}
(A_e)_{ij}=a_0 (Y_e)_{ij},\,(A_d)_{ij}=a_0 (Y_d)_{ij},\,(A_u)_{ij}=a_0 (Y_u)_{ij},
\end{equation}
where $M_i$, $m_{\tilde{\psi}}$, and $A_{\psi}$ are the gaugino mass, the sfermion mass, and the coupling for the soft tri-linear $A$ term, respectively. We first calculate the sfermion mass at the $1~{\rm TeV}$ scale with the SUSY conserving parameters calculated in the last section. For this calculation, we fix $\tan \beta=3$, $(\Lambda_{d_3})_{31}=-0.2$, and $(\Lambda_{d_3})_{32}=0.05$ to realize large contributions to $R_{D^{(\ast)}}$, and $m_{1/2}=900~{\rm GeV}$ to satisfy the current gluino mass limit $M_3 \geq 2~{\rm TeV}$~\cite{mg_RPV_ATLAS, mg_RPC_ATLAS, mg_RPC_CMS}\footnote{This limit depends strongly on the models considered, especially on whether the RPV couplings are present or not. Here we use this as a conservative limit.}.

Moreover, we fix $a_0 = 0$ for the sake of simplicity. The RPV coupling $\Lambda_d$ contributes also to the neutrino masses at the one-loop level, and this contribution would be too large to realize the neutrino mass measurements when the RPV couplings are $\mathcal{O}(1)$ and the sfermion masses are around $\mathcal{O}(1~{\rm TeV})$~\cite{Hall:1983id}. To suppress this one-loop contribution, the relation $a_0 - \mu \tan \beta = 0$ is required in the CMSSM scenario~\cite{Barbier:2004ez}\footnote{Strictly speaking, to realize $\mathcal{O}(1~\mathrm{ eV})$ neutrino masses, $a_0 - \mu \tan \beta < \mathcal{O}(10^{-3})$ GeV when $m_{\tilde{b}} = 1$ TeV and $(\Lambda_{d_3})_{33}=1$. There exist other options to get sub-eV scale neutrino masses, for example, by invoking cancellations between the tri-linear induced contributions and other unrelated contributions to neutrino masses~\cite{Bhattacharyya:1999tv} or by assigning different lepton number charges to the left- and right-handed squarks~\cite{Frugiuele:2011mh}.}, and the parameter $\mu$ should be around the $Z$-boson mass to stabilize the EW scale. Therefore, the parameter $a_0$ is around the EW scale and can be neglected in this calculation\footnote{This assumption is one example for stabilizing the EW scale without fine-tuning. Contrary to this, large and non-negligible $a_0$ can be another choice; for example, to reproduce the 125~GeV Higgs mass~\cite{Higgs_discovery} in the SUSY models, the maximal mixing scenario is useful~\cite{maximal_mixing_scenario}, and in this case $a_0$ should be large and requires unnaturally large fine-tuning to keep $a_0 - \mu \tan \beta$ very small to satisfy neutrino mass constraints. In the case when $a_0$ is small, we should enlarge the particle content of the minimal model by, for example, the stop-like vector-particles that are embedded into ${\bf 10}$ of $SU(5)$, in order to obtain the 125~GeV Higgs mass~\cite{Moroi:1992zk}. Such stop-like vector-particles would contribute to the RG flow for gauge couplings, as discussed in Ref.~\cite{Hisano:2012wq}.}.

\begin{figure}[t]
  \centering
  \includegraphics[width=0.46\textwidth,pagebox=cropbox,clip]{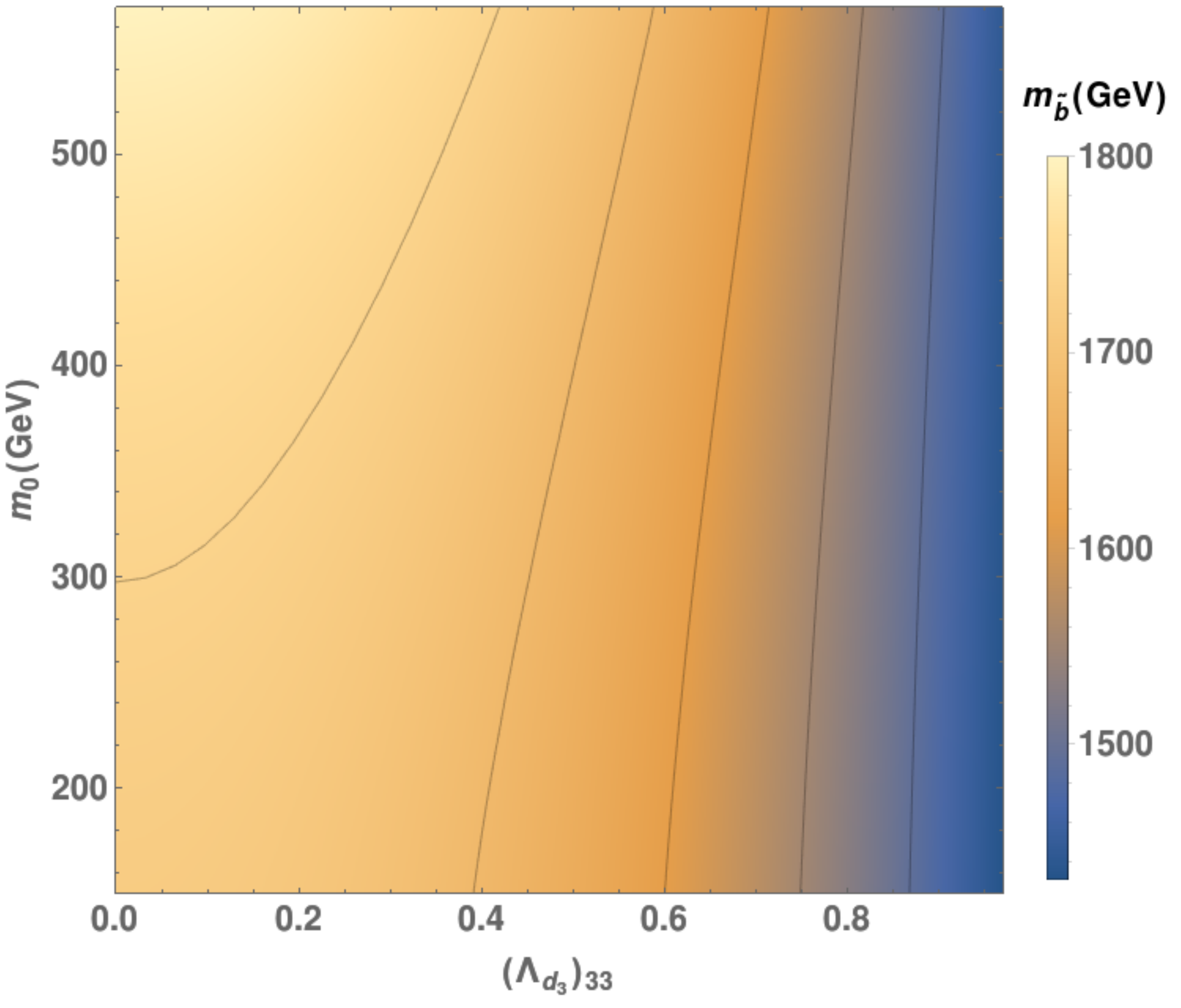}\qquad
  \includegraphics[width=0.46\textwidth,pagebox=cropbox,clip]{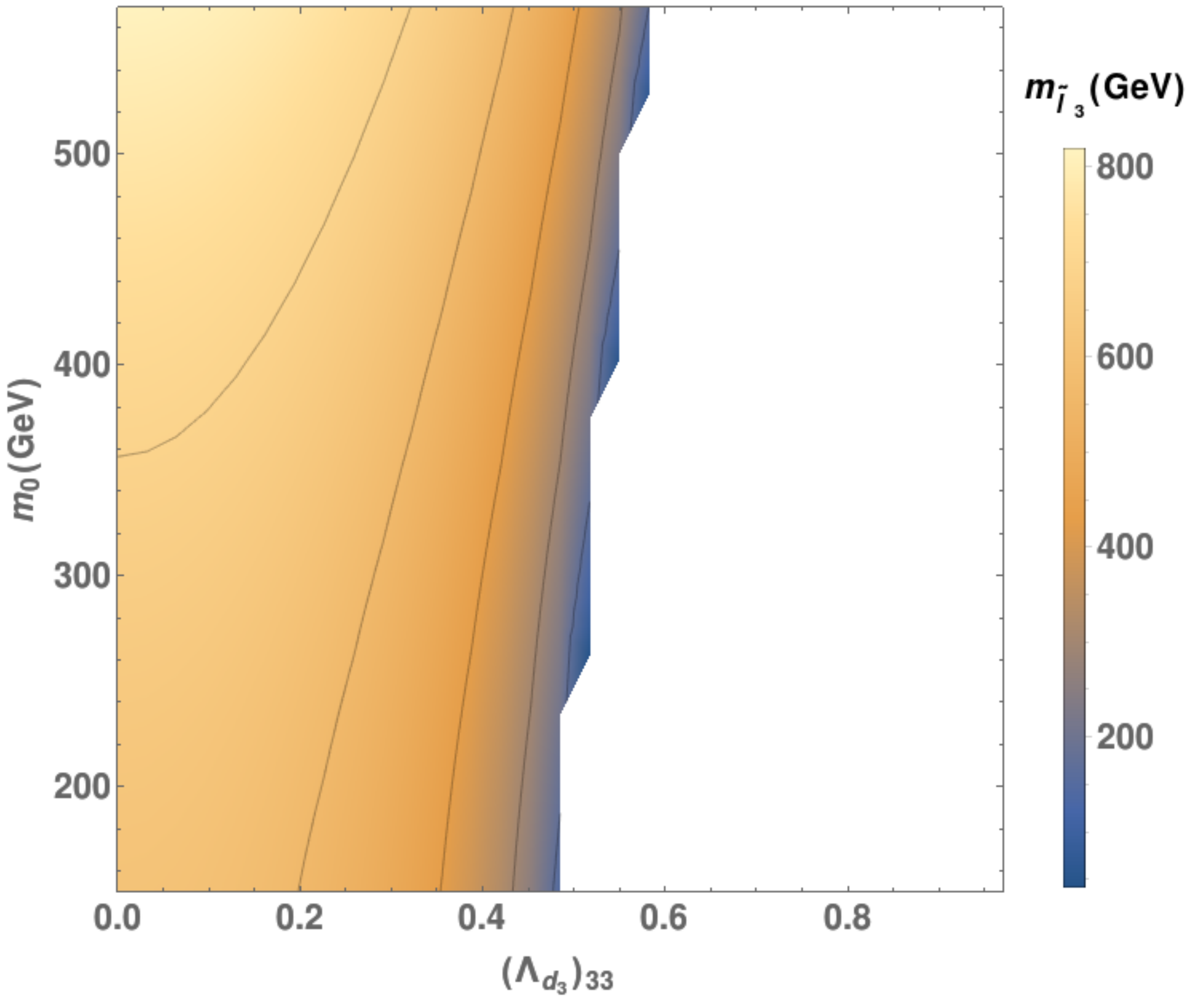}
  \caption{\small Density plots for the sfermion masses $m_{\tilde{b}}$ and $m_{\tilde{l}_{3}}$ at the $1~{\rm TeV}$ scale as a function of the universal sfermion mass $m_0$ and the RPV coupling $(\Lambda_{d_3})_{33}$. Regions ruled out by the presence of tachyons are shown in white.}
\label{fig:Msfermion_univ}
\end{figure}

In Figure \ref{fig:Msfermion_univ}, density plots for the sfermion masses $m_{\tilde{b}}$ and $m_{\tilde{l}_{3}}$ at the $1~{\rm TeV}$ scale are shown as a function of the universal sfermion mass $m_0$ and the RPV coupling $(\Lambda_{d_3})_{33}$.
White region in the right panel is ruled out by a presence of tachyons. From these figures, we can find two problems to realize the $R_{D^{(\ast)}}$ measurements. The first one is the largeness of the sbottom mass: as shown in the left panel of Figure \ref{fig:Msfermion_univ}, the sbottom mass is too heavy to satisfy the $R_{D^{(\ast)}}$ measurements. The second one is the presence of tachyons: as shown in the right panel of Figure \ref{fig:Msfermion_univ}, the slepton becomes tachyonic in the large RPV coupling region. Therefore, to accommodate the $R_{D^{(\ast)}}$ anomaly, these two problems have to be resolved in the CMSSM scenario.

The gaugino contributions to the $\beta$-functions for the sfermion masses play an important role in understanding these problems. In the CMSSM, the squark masses are larger than the slepton masses in the low-energy scale. This feature is due to the gaugino contributions. While these contributions make sfermion masses large in the low-energy scale, their effects on the slepton masses are relatively small because of the absence of the gluino contribution,  which is the largest among the gaugino contributions. As the gluino mass should be larger than $2~{\rm TeV}$~\cite{mg_RPV_ATLAS, mg_RPC_ATLAS, mg_RPC_CMS}, the gluino contribution is large, making the right-handed sbottom mass becomes larger than $1~{\rm TeV}$. On the other hand, the RPV contributions make the slepton mass small, and overcome the small gaugino contributions, making therefore the squared slepton masses negative.

\subsection{Non-universal sfermion masses \label{sec.3.1}}

\begin{figure}[t]
  \begin{tabular}{cc}
  \begin{minipage}{0.5\hsize}
  \includegraphics[width=0.9\textwidth,pagebox=cropbox,clip]{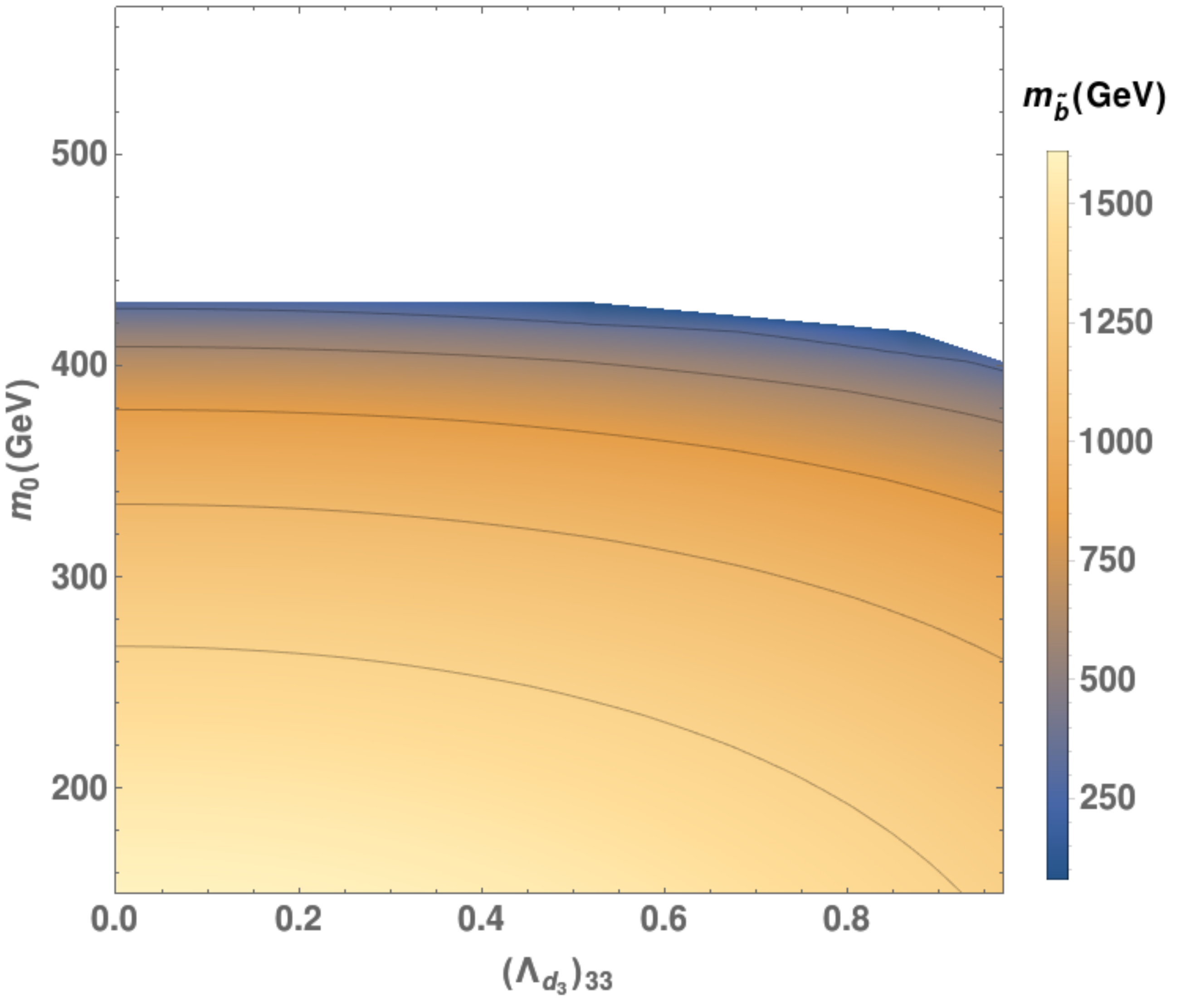}
  \centering
  {\small non-universal sfermion masses}
  \end{minipage}&
  \begin{minipage}{0.5\hsize}
  \includegraphics[width=0.9\textwidth,pagebox=cropbox,clip]{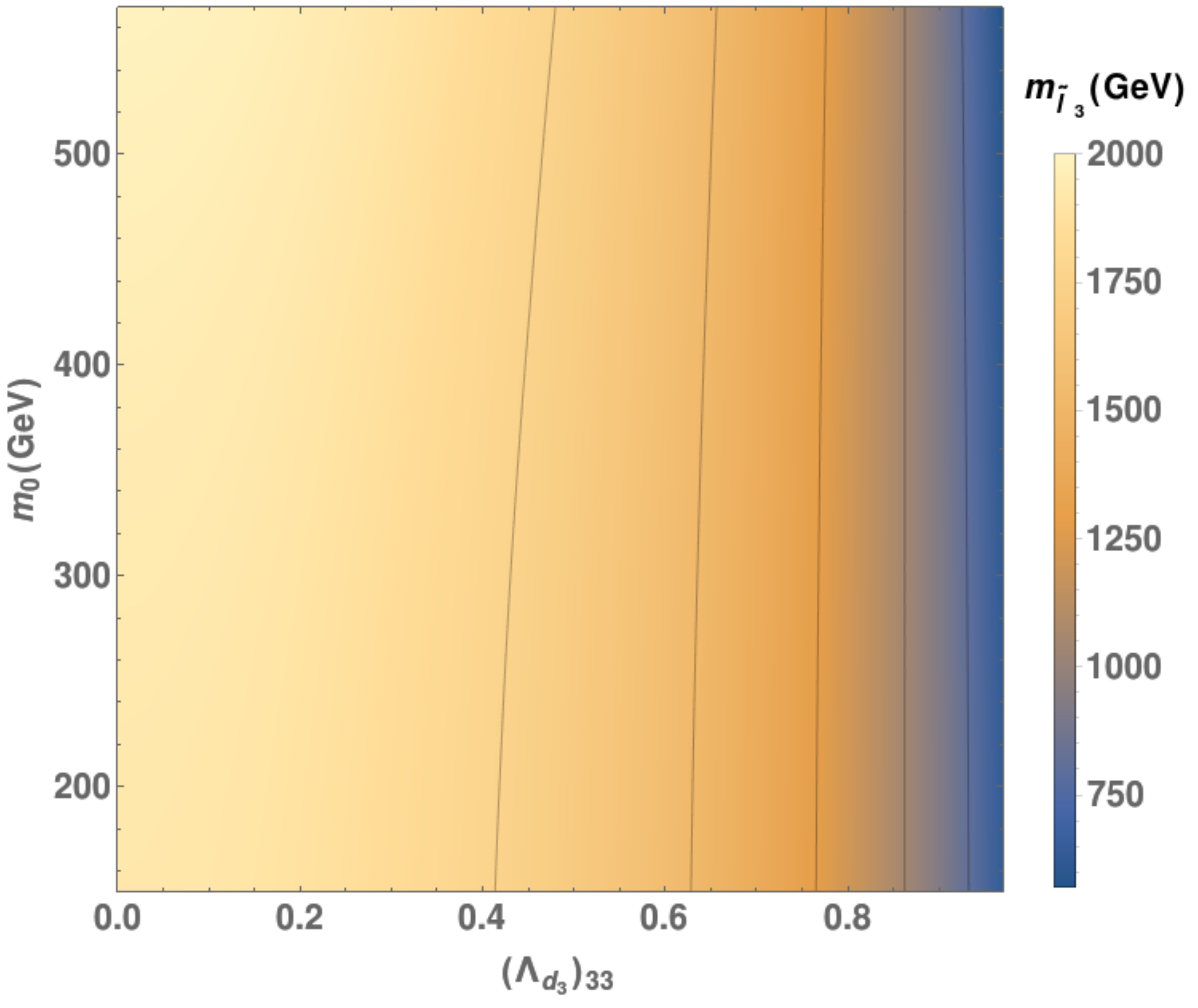}
  \centering
  {\small non-universal gaugino masses}
  \end{minipage}
\end{tabular}
  \caption{\small The captions are the same as those in Figure \ref{fig:Msfermion_univ}, but now with non-universal sfermion masses (left) and non-universal gaugino masses (right).}
\label{fig:Msfermion_non}
\end{figure}

Non-universal sfermion masses are helpful for resolving the first problem, \textit{i.e.} the largeness of the sbottom mass. In this paper, the non-universal sfermion masses are specified as
\begin{equation}
m_{\tilde{\psi}}^2 = \left(
\begin{array}{ccc}
m_{12}^2 & 0 & 0 \\
0 & m_{12}^2 & 0 \\
0 & 0 & m_0^2 \\
\end{array}
\right),\quad
m_{12} \equiv r_m m_0.
\end{equation}
When $r_m \gg 1$, the natural SUSY scenario is realized, where the third generation sfermion masses become much lighter than that of the first two generations. With the two-loop $\beta$-functions for the sfermion masses~\cite{Martin:1993zk}, the resulting sfermion masses, especially the squark masses, become small. In particular, contributions from the trace term $\text{Tr}\, m_{\tilde{\psi}}^2=(1+2r_m^2)m_0^2$ have a much stronger effect on the third-generation squark masses. In the left panel of Figure \ref{fig:Msfermion_non}, the density plots for $m_{\tilde{b}}$ at the $1~{\rm TeV}$ scale are shown as a function of the sfermion mass $m_0$ and the RPV coupling $(\Lambda_{d_3})_{33}$, when $r_m = 20$. It can be seen that, owing to the non-universality of sfermion masses, the right-handed sbottom mass can now be around $1~{\rm TeV}$, together with $\mathcal{O}(1)$ RPV couplings.

\subsection{Non-universal gaugino masses \label{sec.3.2}}

The gaugino contributions to the slepton masses are enlarged when the bino mass becomes large. In the GUT models, while universal gaugino masses are usually assumed, non-universal gaugino masses are also feasible~\cite{non_universal_gaugino}. In the $SU(5)$ GUT models, the tensor product of the adjoint fields is decomposed as \begin{equation}
(\bm{24} \otimes \bm{24})_{\text{sym}} = \bm{1} \oplus \bm{24} \oplus \bm{75} \oplus \bm{200}.
\end{equation}
In the CMSSM, we assume that this tensor product becomes singlet. However, when it is not singlet, non-universal gaugino masses are achieved by the $F$-term breaking of the VEVs of the non-singlet scalar fields. When the $F$-term breaking is assumed, the VEV of the dimensional-$75$ scalar field would induce nonzero gaugino masses, with the mass relation at the GUT scale given by
\begin{equation}
M_1 : M_2 : M_3 = -5 : 3 : 1.
\end{equation}
Therefore, the bino mass can be large. The same observation is also applied to the case with the dimensional-$200$ scalar field. In the right panel of Figure \ref{fig:Msfermion_non}, the density plots for $m_{\tilde{l}_3}$ at the $1~{\rm TeV}$ scale are shown as a function of the universal sfermion mass $m_0$ and the RPV coupling $(\Lambda_{d_3})_{33}$, when $M_1 : M_2 : M_3 = -5 : 3 : 1$ at the GUT scale. Owing to the non-universality of gaugino masses, the tachyonic region disappear. Therefore, this non-universality is helpful for resolving the second problem, \textit{i.e.} the presence of tachyons.

\subsection{Non-universal sfermion and gaugino masses \label{sec.3.3}}

In the subsections \ref{sec.3.1} and \ref{sec.3.2}, we have shown that the non-universal sfermion and gaugino masses are helpful for resolving the two problems encountered in the CMSSM scenario, to realize the $R_{D^{(\ast)}}$ measurements. In this subsection, we discuss the sfermion masses when these two non-universalities are assumed.

\begin{figure}[t]
  \centering
  \includegraphics[width=0.46\textwidth,pagebox=cropbox,clip]{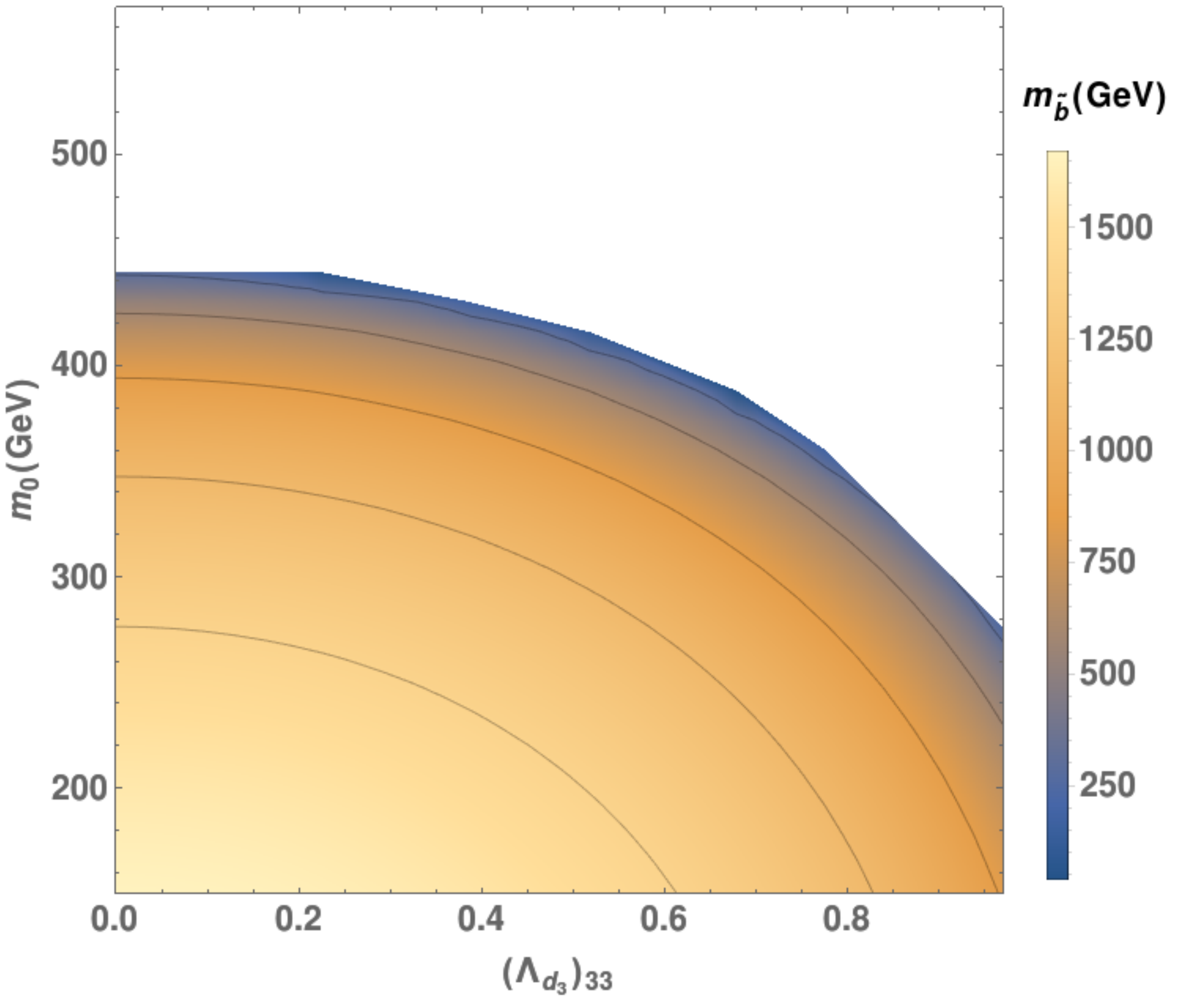}\qquad
  \includegraphics[width=0.46\textwidth,pagebox=cropbox,clip]{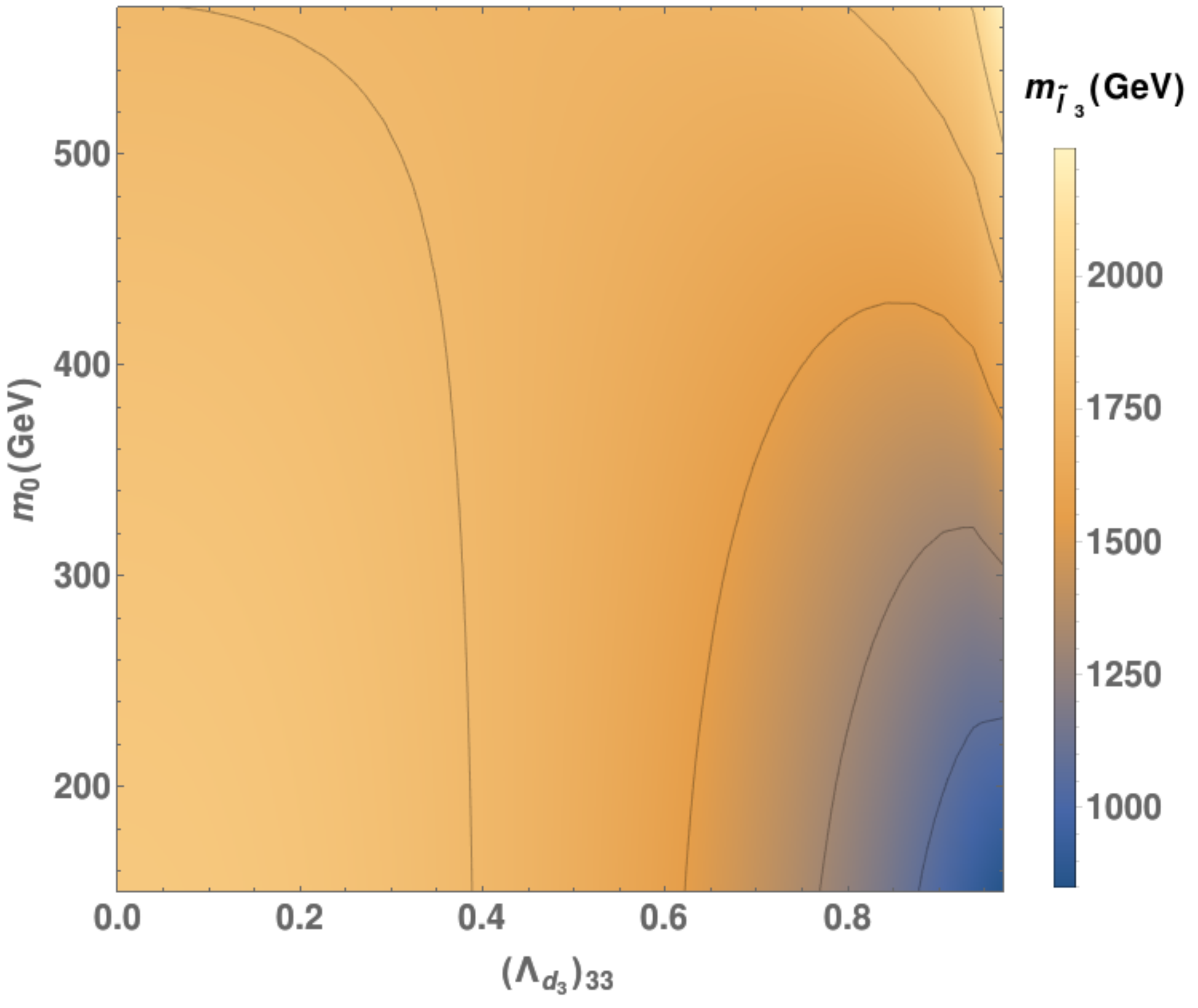}
  \caption{\small The captions are the same as those in Figure \ref{fig:Msfermion_univ}, but now with non-universal sfermion and gaugino masses simultaneously.}
\label{fig:Msfermion_double}
\end{figure}

In Figure \ref{fig:Msfermion_double}, density plots for $m_{\tilde{b}}$ and $m_{\tilde{l}_3}$ at the $1~{\rm TeV}$ scale are shown as a function of the sfermion mass $m_0$ and the RPV coupling $(\Lambda_{d_3})_{33}$. For this calculation, we assume $r_m = 20$ and $M_1 : M_2 : M_3 = -5 : 3 : 1$ at the GUT scale. From this figure, we can see that, thanks to these two non-universalities, large RPV couplings and around $1~{\rm TeV}$ right-handed sbottom mass can be realized simultaneously. In Appendix \ref{app.1}, density plots for the other sfermion masses $m_{\tilde{q}_3}$, $m_{\tilde{t}}$, $m_{\tilde{\tau}}$ at the $1~{\rm TeV}$ scale are also shown.

From the above discussions, we can conclude model-independently that, among the three tri-linear RPV couplings, $\Lambda_d$ must be large enough to reproduce the ratio $R_{D^{(*)}}$ and, at the same time, $\Lambda_u$ must be zero in order to realize the proton decay constraint. This leaves with us two remaining choices for $\Lambda_e$, being either zero or sizable. Up to now our discussions are restricted to the case with zero $\Lambda_e$. The choice with sizable $\Lambda_e$ can be realized through the matter-Higgs mixing mechanism~\cite{Smirnov:1995ey} and will also affect the RG flows for gauge couplings, Yukawa couplings and squared sfermion masses. Especially, large $\Lambda_e$ would make the squared sfermion masses for the left-handed lepton doublets negative, in the same way as large $\Lambda_d$ does for the right-handed sleptons. Such a problem can, however, be solved by the non-universal gaugino mass too.

\section{Discussion and summary \label{sec.4}}

In this paper, motivated by the excess in the $R_{D^{(\ast)}}$ measurements by the $B$-physics experiments, we have investigated the RPV interactions for resolving the $R_{D^{(\ast)}}$ anomaly with the GUT assumption.

It has been found that, to resolve the $R_{D^{(\ast)}}$ anomaly, $\mathcal{O}(1)$ RPV coupling and around $1~{\rm TeV}$ sbottom mass are required. It has also been shown that this large RPV coupling is conducive to realize the bottom-tau Yukawa unification that appears in the $SU(5)$ GUT models and, to realize this unification, small $\tan \beta$ is particularly favored. On the other hand, problems appear for realizing the favorable sfermion masses: large RPV couplings would make the sleptons tachyonic, and around $1~{\rm TeV}$ sbottom mass is not acceptable when the CMSSM relations for the SUSY parameters are assumed. For solving these problems, two non-universalities, the non-universal sfermion masses and the  non-universal gaugino masses, have been found especially favorable. These non-universalities are also motivated for solving the EW-scale stabilization problem, and can be naturally introduced within the GUT assumption.

The non-universalities of sfermion and gaugino masses can be examined in the future; especially when the third-generation squark masses are around $1~{\rm TeV}$, chromo-electric dipole moments can be a strong signal candidate, although they depend strongly on the quark mixings~\cite{Maekawa:2017xci}. Another consequence for our scenario is the relation for suppressing the dangerous one-loop contribution to the neutrino masses, $a_0 - \mu \tan \beta = 0$. This suppression means that the mass insertion parameter $\delta^d_{LR}$ is zero and, therefore, the SUSY flavor violating contribution from this mass insertion parameter is strongly suppressed. These topics are beyond the scope of this paper and will be explored in the future.

\section*{Acknowledgments}

This work is supported by the National Natural Science Foundation of China under Grant Nos.~11675061, 11775092, 11521064 and 11435003.
X.L. is also supported in part by the self-determined research funds of CCNU from the colleges' basic research and operation of MOE~(CCNU18TS029). Q.H. is also supported by the China Postdoctoral Science Foundation (2018M632896).
Y.M. is also supported by the International Postdoctoral Exchange Fellowship Program (IPEFP).

\appendix

\section{Relevant formulae for the processes used to produce Figure \ref{fig:RPV_para_region} \label{app.eq}}

As we find that there is a global factor $1/2$~($-1/2$) missing for the NP terms in Eqs.~(9) and (12)~(Eqs.~(20) and (21)) of Ref.~\cite{Altmannshofer:2017poe}, which we are closely following, we decide to give in this appendix all the relevant formulae for the processes used to produce Figure \ref{fig:RPV_para_region}.

At low energies, the effective Lagrangian for $d_j \to u_n \tau \nu_\tau$ transition is given by
\begin{equation}\label{eq:Lefflow}
{\cal L}_\text{eff}= - \frac{4G_F}{\sqrt{2}} (V_{CKM})_{nj} (1+C_{V_L}^{nj}) (\bar{u}_n \gamma_\mu P_L d_j)(\bar{\tau}\gamma^\mu P_L \nu_\tau)+ {\rm h.c.},
\end{equation}
where $C_{V_L}^{nj}$ encodes all the contribution from the RPV couplings.
Matching Eq.~\eqref{eq:Leff} onto Eq.~\eqref{eq:Lefflow}, one gets
\begin{equation}
C_{V_L}^{nj} = \frac{v^2(\Lambda_{d_3})_{3j}}{4m^2_{\tilde{b}}} \sum_{n'}(\Lambda_{d_3})_{3n'}^\ast\frac{(V_{CKM})_{nn'}}{(V_{CKM})_{nj}},
\end{equation}
with $v=246$ GeV being the Higgs VEV. The branching ratio of the decay governed by $d_j \to u_n \tau \nu_\tau$ transition can then be written as ${\cal B}/{\cal B}_\text{SM}=|1+C_{V_L}^{nj}|^2$; explicitly, we find
\begin{align}
\frac{R_D}{R_D^\text{SM}}=\frac{R_{D^\ast}}{R_{D^\ast}^\text{SM}} & =|1+C_{V_L}^{23}|^2,
&
\frac{{\cal B}(B\to \tau\nu)}{{\cal B}(B\to \tau\nu)_\text{SM}}& =|1+C_{V_L}^{13}|^2,
\\[2mm]
\frac{{\cal B}(\tau\to \pi\nu)}{{\cal B}(\tau\to \pi\nu)_\text{SM}}& =|1+C_{V_L}^{11}|^2,
&
\frac{{\cal B}(\tau\to K\nu)}{{\cal B}(\tau\to K\nu)_\text{SM}}& =|1+C_{V_L}^{12}|^2,
\end{align}
all of which have to satisfy the constraints summarized in Table \ref{tab:const_diff}.

Using ${\cal B}(\pi^-\to\mu^-\nu)=99.9877\%$ and ${\cal B}(K^-\to\mu^-\nu)=(63.56\pm0.11)\%$~\cite{Patrignani:2016xqp}, as well as~\cite{Pich:2013lsa}
\begin{equation}\label{eq:tauraredecay}
{\cal B}(\tau^-\to P^-\nu)_\text{SM}=\frac{\tau_\tau m_\tau^3(1-m_P^2/m_\tau^2)^2}{2\tau_P m_P m_\mu^2(1-m_\mu^2/m_P^2)^2}(1+\delta R_{\tau/P}){\cal B}(P^-\to\mu^-\nu),
\end{equation}
where $\tau_\tau$ and $\tau_P$ are the lifetime of the $\tau$ lepton and the $P$ meson, respectively, we can obtain the constraints on the RPV couplings from these decays. The factor $\delta R_{\tau/P}$ in Eq.~\eqref{eq:tauraredecay} denotes the relative radiative correction and has been estimated to be $\delta R_{\tau/\pi}=(0.16\pm0.14)\%$ and $\delta R_{\tau/K}=(0.90\pm0.22)\%$~\cite{Pich:2013lsa,Marciano:1993sh}. The SM prediction for the branching fractions of $\tau\to \pi\nu$ and $\tau\to K\nu$ are shown in Table \ref{tab:const_diff}.

The formulas for other processes are given, respectively, by
\begin{align}
 \frac{\mathcal{B}(B \to K \nu\nu)}{\mathcal{B}(B \to K\nu\nu)_\text{SM}} &= \frac{2}{3} + \frac{1}{3} \left| 1 - \frac{v^2}{2m^2_{\tilde b}} \frac{\pi \sin^2 \theta_W}{\alpha X_t} \frac{(\Lambda_{d_3})_{33} (\Lambda_{d_3})_{32}^*}{(V_{CKM})_{33} (V_{CKM})_{32}^*} \right|^2,\\[0.2cm]
 \frac{\mathcal{B}(B \to \pi \nu\nu)}{\mathcal{B}(B \to \pi\nu\nu)_\text{SM}} &= \frac{2}{3} + \frac{1}{3} \left| 1 - \frac{v^2}{2m^2_{\tilde b}} \frac{\pi \sin^2 \theta_W}{\alpha X_t} \frac{(\Lambda_{d_3})_{33} (\Lambda_{d_3})_{31}^*}{(V_{CKM})_{33} (V_{CKM})_{31}^*}  \right|^2,
\end{align}
with the SM loop function $X_t=1.469\pm0.017$~\cite{Buras:2014fpa}. For the corrections to the $Z$ and $W$ couplings to leptons, we have
\begin{align}
 \frac{g_{Z\tau_L\tau_L}}{g_{Z l_L l_L}} &= 1 - \frac{3 |(\Lambda_{d_3})_{33}|^2}{16\pi^2} \frac{1}{1 - 2 \sin^2 \theta_W} \frac{M_t^2}{m^2_{\tilde b}} f_Z\left(\frac{M_t^2}{m^2_{\tilde b}}\right), \\[0.2cm]
 \frac{g_{W\tau_L\nu_\tau}}{g_{W l_L\nu_ l}} &= 1 - \frac{3 |(\Lambda_{d_3})_{33}|^2}{16\pi^2} \frac{1}{4} \frac{M_t^2}{m^2_{\tilde b}} f_W\left(\frac{M_t^2}{m^2_{\tilde b}}\right),
\end{align}
where the loop functions are given by $f_Z(x) = \frac{1}{x-1} - \frac{\log(x)}{(x-1)^2}$, $f_W(x) = \frac{1}{x-1} - \frac{(2-x)\log(x)}{(x-1)^2}$~\cite{Feruglio:2016gvd}.

\section{Sfermion masses in the case with non-universal sfermion and gaugino masses \label{app.1}}

\begin{figure}[t]
  \centering
  \includegraphics[width=0.45\textwidth,pagebox=cropbox,clip]{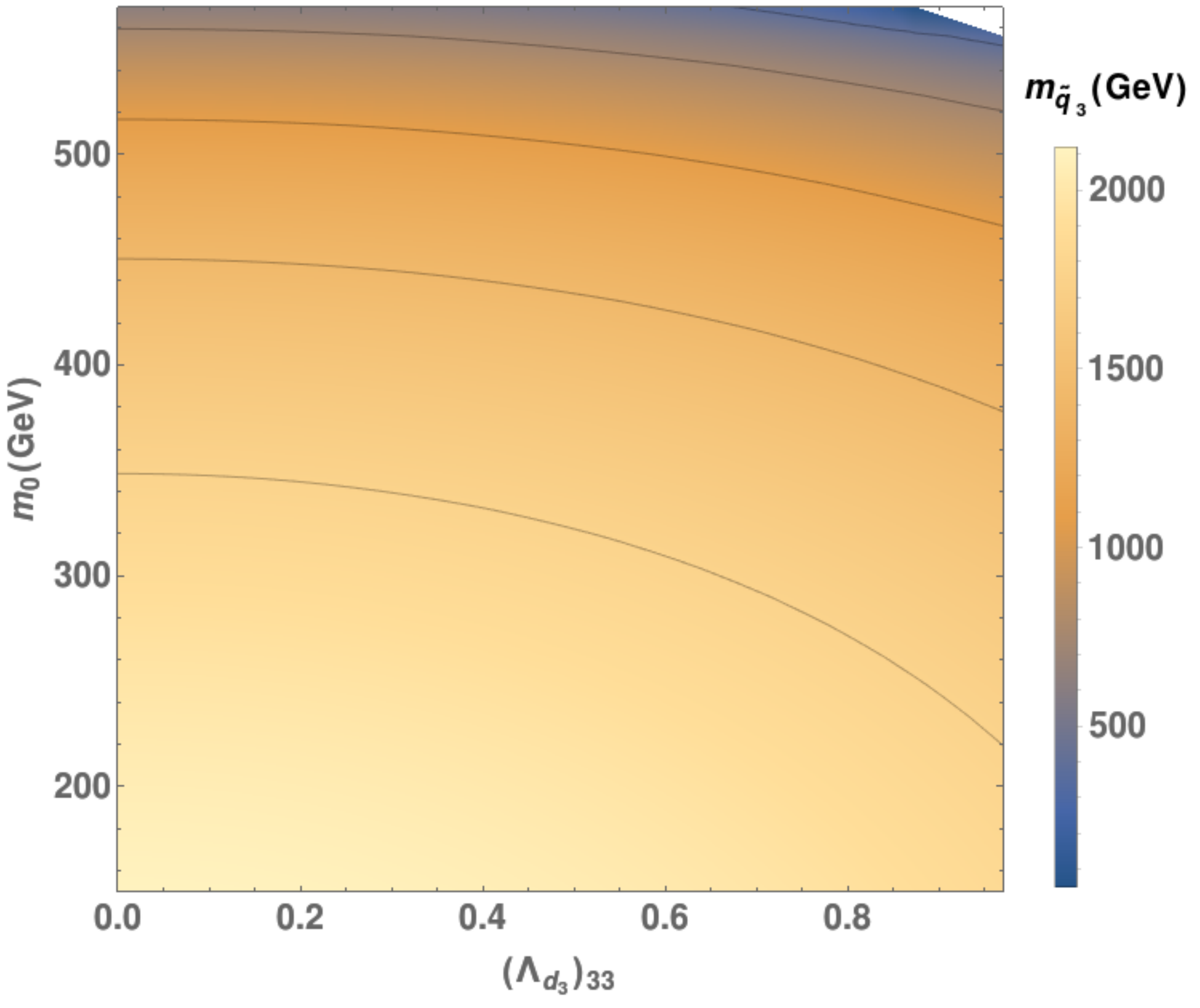}\qquad
  \includegraphics[width=0.45\textwidth,pagebox=cropbox,clip]{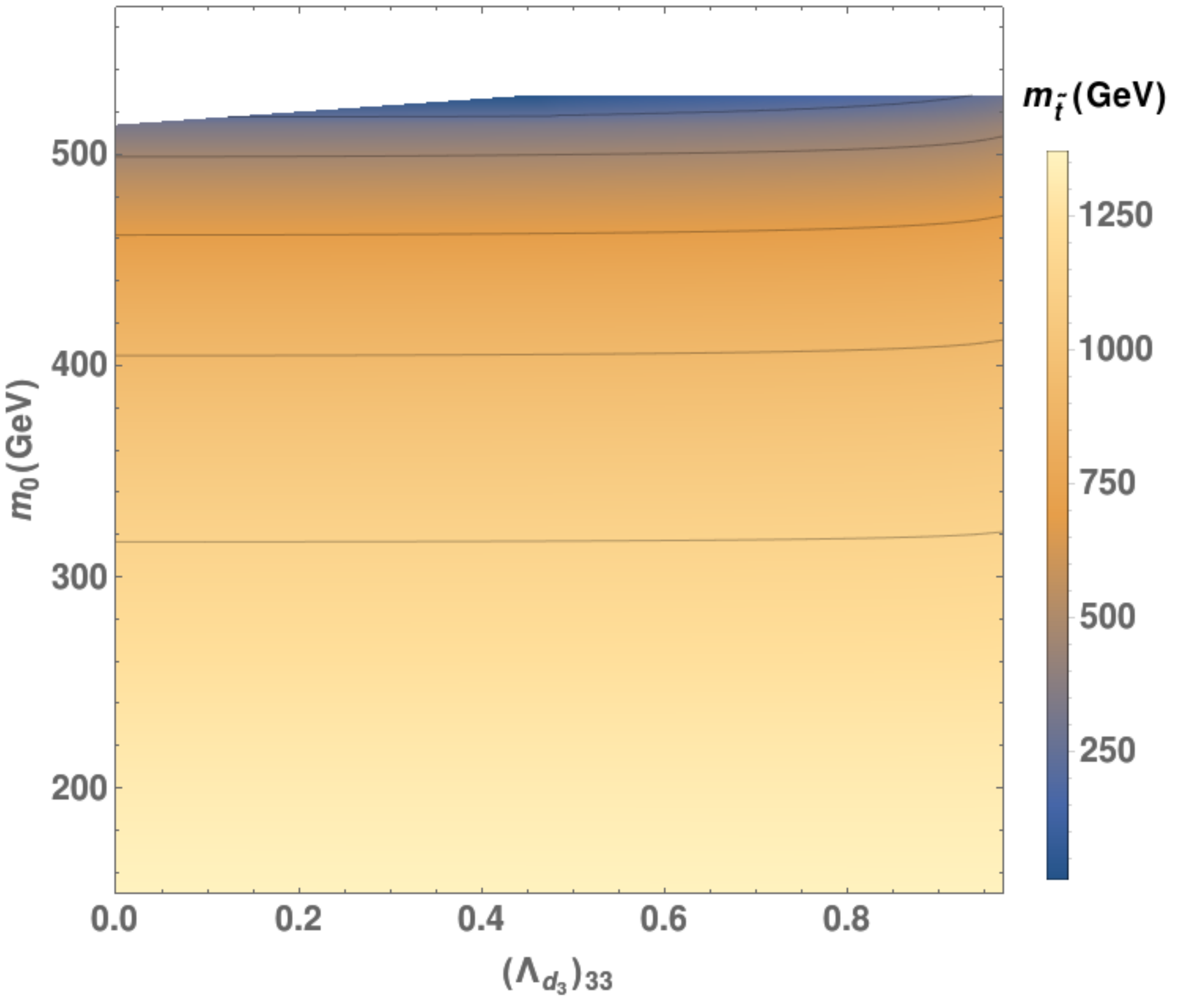}\\[0.4cm]
  \includegraphics[width=0.45\textwidth,pagebox=cropbox,clip]{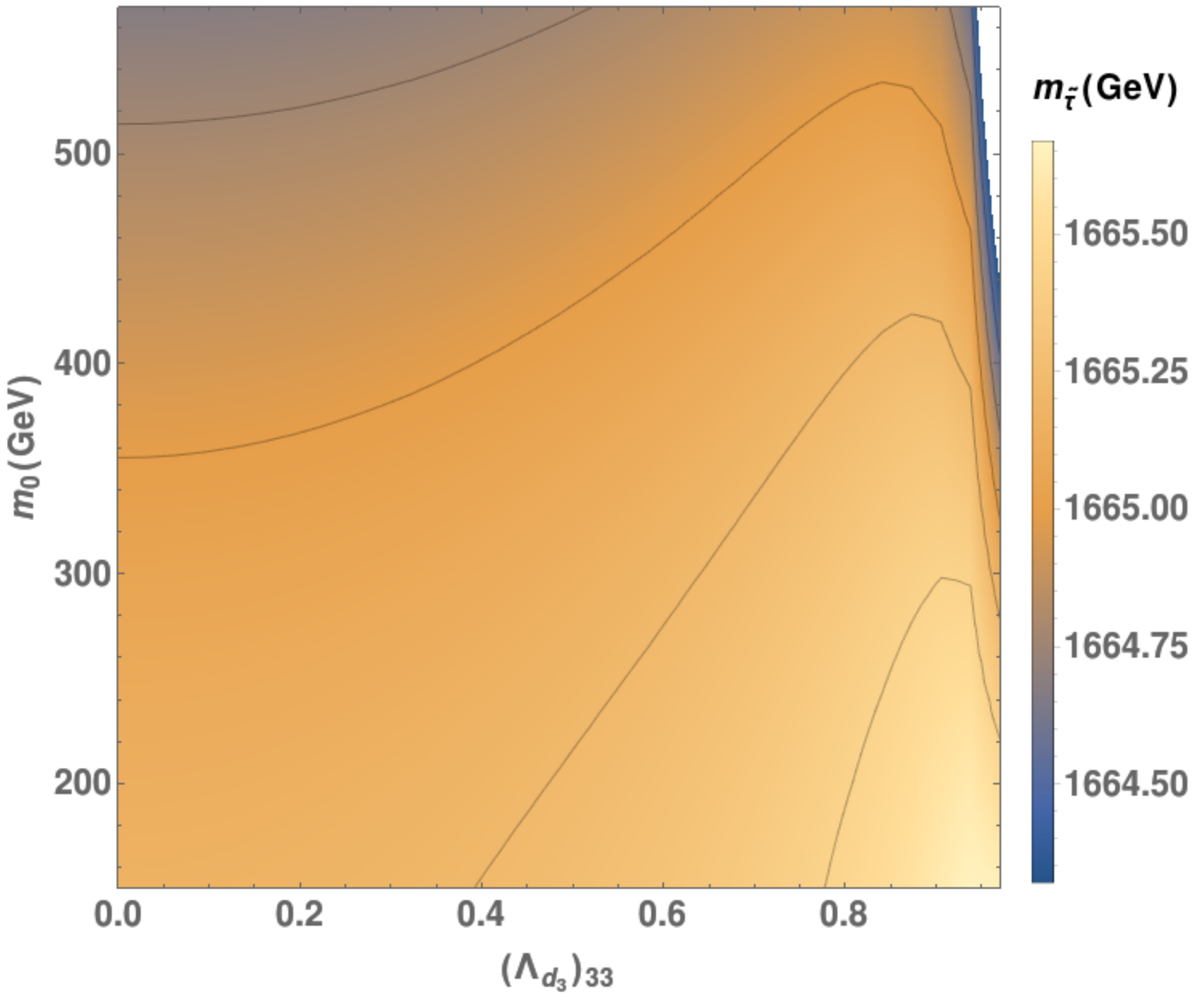}
  \caption{\small Density plots for the sfermion masses $m_{\tilde{q}_{3}}$, $m_{\tilde{t}}$, and $m_{\tilde{\tau}}$ at the $1~{\rm TeV}$ as a function of the sfermion mass $m_0$ and the RPV coupling $(\Lambda_{d_3})_{33}$. Regions ruled out by the presence of tachyons are shown in white.}
\label{fig:Msfermion_other}
\end{figure}

In Figure \ref{fig:Msfermion_other}, density plots for $m_{\tilde{q}_{3}}$, $m_{\tilde{t}}$, and $m_{\tilde{\tau}}$ at $1~{\rm TeV}$ are shown as a function of the sfermion mass $m_0$ and the RPV coupling $(\Lambda_{d_3})_{33}$. For this calculation, $r_m = 20$ and $M_1 : M_2 : M_3 = -5 : 3 : 1$ at the GUT scale have been assumed.

From these plots, we can see that, with the simultaneous presence of non-universal sfermion and gaugino masses, parameter regions with large RPV couplings and around $1~{\rm TeV}$ sbottom mass, required for resolving the $R_{D^{(\ast)}}$ anomaly, can be naturally obtained in our scenario, avoiding at the same time the presence of any tachyons.

\end{document}